# Forecasting the effect of heat stress index and climate change on cloud data center energy consumption

Vikas Ramachandra
Adjunct Professor, Data Science
University of California, Berkeley
Berkeley, CA
virama@berkeley.edu

## Abstract

In this paper, we estimate the effect of heat stress index (a measure which takes into account rising temperatures as well as humidity) on data center energy consumption. We use forecasting models to predict future energy use by data centers, taking into account rising temperature scenarios. We compare those estimates with baseline forecasted energy consumption (without heat stress index or rising temperature correction) and present the result that there is a sizeable and significant difference in the two forecasts. We show that rising temperatures will cause a negative impact on data center energy consumption, increasing it by about 8 percent, and conclude that data center energy consumption analyses and forecasts must include the effects of heat stress index and rising temperatures and other climate change related effects.

## 1.Introduction

In 2014, data centers in the U.S. consumed an estimated 70 billion kWh, representing about 1.8% of total U.S. electricity consumption. Current study results show data center electricity consumption increased by about 4% from 2010-2014 [1,2]. Energy use is expected to continue slightly increasing in the near future, increasing 4% from 2014-2020, the same rate as the past five years. Based on current trend estimates, U.S. data centers are projected to consume approximately 73 billion kWh in 2020. About 33% of cloud computing energy goes into cooling [3]. However, these trend estimates are based on extrapolating present usage (i.e. forecasting into the future) under the 'business as usual' scenario.

Earlier research work towards estimation of data center energy consumption does not take the effect of rising temperatures and other climate change related events into account. In this paper, we incorporate the impact of climate change on data center performance. For example, one of the major effects of higher temperatures and humidity will be additional stress on the cooling equipment and energy required to cool data centers. Here, we quantify the heat stress related impact on energy requirements of data centers, and forecast those requirements into the future, taking into account different climate change scenarios as well.

The research question we address is as follows: *Is there a significant increase in data center energy requirements when global heat stress index increases?*
In other words, what is the impact of climate change as well as rising temperatures on data centers. In the next section, we present analysis and simple forecasting models to answer this question.

## 2.Forecasting data center energy consumption under rising temperatures

**Methodology Summary**

The most accurate way to measure heat stress impact on data center energy consumption is by global temperature increase, which can be calculated through forecasted $CO_2$ emissions. $CO_2$ emission forecasts vary from moderate to vigorous scenarios (RCPs). In this test, we use RCP 4.5(moderate) and RCP 8.5 (vigorous) as test cases [7]. The IPCC AR5 $CO_2$ emission forecasts was used as a source determine $CO_2$ levels from 2000-2050. Change in global temperature is then calculated from the IPCC equation, converted into degrees Fahrenheit. According to earlier research [4], an increase in data center temperature by 1 degree Fahrenheit will increase energy consumption by 2-8%. In this paper, we use the mean impact on energy consumption, 5%. The increase is temperature is then converted into a percent increase in energy required for the data centers. There are 2 variable increases (1. Under RCP 4.5, 2. Under RCP 8.5). These increases are applied to existing energy consumption forecasts, used from Data Center Energy Usage Report from Berkeley National Laboratory [2]. Two scenarios are used (1. 2016 Current Trend 2. Best Practices Scenarios with improved energy efficiency practices considered). For each scenario, the RCP 4.5 and RCP 8.5 trends are forecast, resulting in a total of 4 separate trend lines showing the impact of global temperature increase on energy consumption of U.S. data centers. The 2 control scenarios, for which temperature changes are not accounted for, are also modeled to provide a baseline expectation. These are plotted in the figure below, the discussion below explains various lines in the plot.

The assumptions made are as follows:
- Global temperature rise of 2.5 to 10 degrees F over the next century (for different scenarios)
- Temperatures will increase annually on a linear basis
- Moderate emissions are based on IPCC RCP 4.5 $CO_2$ forecasts
- Vigorous emissions are based on IPCC RCP 8.5 $CO_2$ forecasts

- Change of 1 degree Fahrenheit will increase energy consumption cost by 5%
- After 2020, energy efficiency standards will reach asymptotic improvement in energy efficiency

This leads to the following scenarios (as can be seen in the figure below)
1.a: Moderate emissions annual outlook (according to IPCC) & energy efficiencies reduce consumption by 40% until 2020 — Brown dotted Line — *Best case*
1.b: Moderate emissions annual outlook & energy efficiencies remain the same — Red dashed Line
2.a: Vigorous emissions annual outlook & energy efficiencies reduce consumption by 40% until 2020 -- Green solid Line
2.b: Vigorous emissions annual outlook & energy efficiencies remain the same — Blue dashed Line — *Worst case*

We can compare the energy consumption considering estimated temperature rise scenarios with their respective control plot (controls do not take climate change into account). For the dashed blue line (vigorous emissions based temperature effects and energy consumption forecast) and solid red line (moderate emissions based temperature effects and energy consumption forecast), the dotted black line is the control (climate change effects ignored on energy consumption). Thus, comparing the difference between the blue-black lines and the red-black lines gives us the additional energy needed by data centers (8%), taking into account rising temperatures and other climate change effects.

Similarly, we can compare the energy consumption forecast with the control (no climate change effect corrections), in the presence of improved energy efficiency practices for data centers.
The dashed yellow line (vigorous emissions based temperature effects and energy consumption forecast) and solid green line (moderate emissions based temperature effects and energy consumption forecast), the dotted brown line is the control (climate change effects ignored on energy consumption), all in the presence of improved energy efficiency practices.
Note that one can observe that for this group of lines, there is a dip in the energy consumption for some years, followed by an increase thereafter. The reason for this is as follows: it is expected that for the next few years, efficient energy practices for data centers will actually reduce the energy consumption. However, once maximum efficiency is reached, the consumption levels will start to increase again, This is independent of the effects of climate change. In this case too, rising temperatures will mean that data centers require about 8% additional energy for operation.

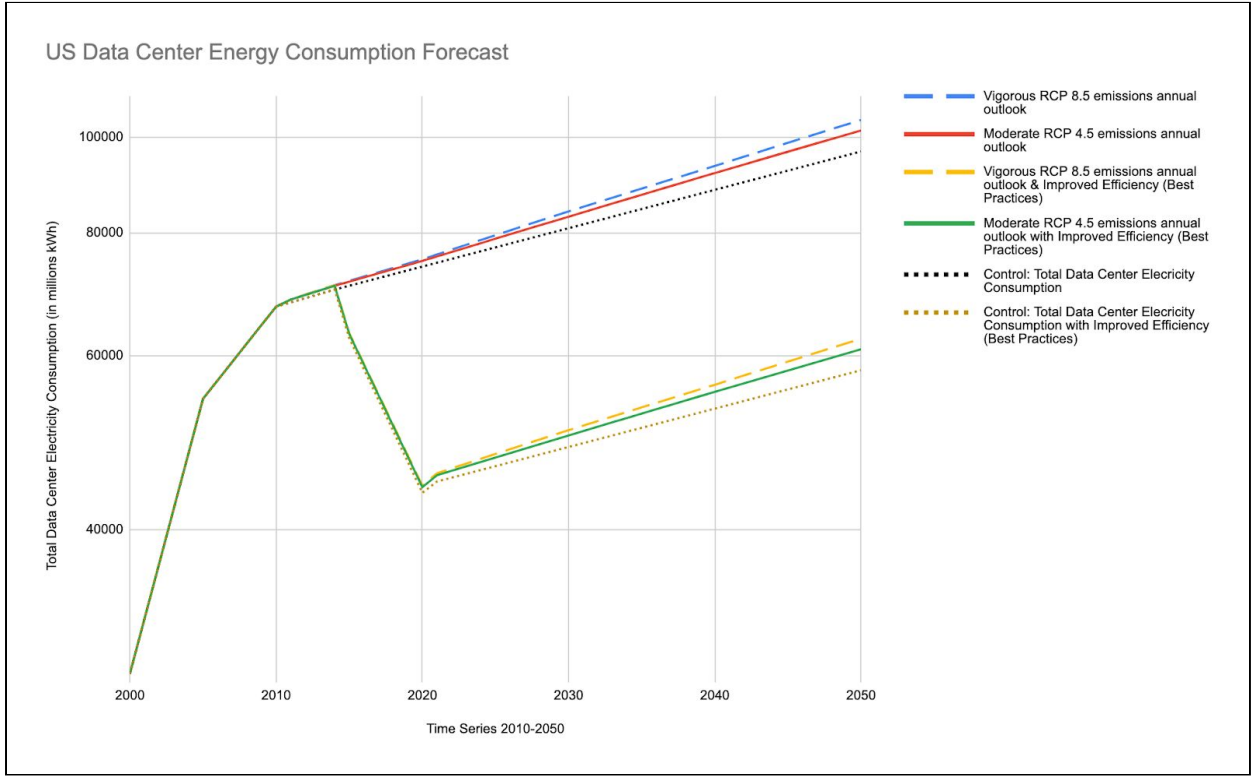

Figure: Energy consumption for various climate scenarios (in million Kwh), forecast by year

**Data sources and modeling equations**

The data source for the IPCC climate scenarios is from [5,6,7]. The data center energy consumption dataset comes from [1,2]. We follow these simple 3 steps to get the forecasts (repeated for various scenarios). First, we use the IPCC carbon dioxide levels, and relation to temperature change for the future. Next, we convert the temperature to Fahrenheit, and use a linear model with slope 0.05 (5%) which relates energy consumption increase to temperature (degree F) increase.

Step 1. IPCC $CO_2$ Equation:

*Change in Temp(Celsius) = 1.66 * ln(Carbon(Year)/Carbon(Year 0--Baseline))*

Step 2. Convert to Fahrenheit:

*Change in Temp(Fahrenheit) = 1.8 * (Change in Temp(Celsius))*

Step 3. Linear model: 5% slope/increase in energy consumption for every +1 degree Fahrenheit:

*Energy(new) = Energy(Old) * (1 + (0.05 * Change in Temp(Fahrenheit)))*

# 3. Conclusion

In this paper, we have incorporated the effect of rising temperatures on data center energy consumption calculations. We have compared various climate change scenarios and estimated the increase in energy requirements. A simple linear model based forecast shows that data center energy consumption will increase by 8% due to rising temperatures alone, by 2050, as compared to forecasts which do not take climate change factors into account. To the best of our knowledge, this is the first attempt to include the impact of rising temperatures on data center energy consumption requirements and forecasts. Future work will include adding the effects of other climate related variables to the data center energy requirements forecasts.